\documentclass[twocolumn]{aastex62}
\usepackage{amsmath}
\usepackage{graphicx}

\usepackage{txfonts}

\usepackage{color, colortbl}
\usepackage{natbib}
\colorlet{RED}{red}

\bibpunct{(}{)}{;}{a}{}{,}

\def\setsymbol#1#2{\expandafter\def\csname #1\endcsname{#2}}
\def\getsymbol#1{\csname #1\endcsname}

\newbox\tablebox    \newdimen\tablewidth
\def\leaderfil{\leaders\hbox to 5pt{\hss.\hss}\hfil}
\def\endPlancktablewide{\tablewidth=\textwidth 
    $$\hss\copy\tablebox\hss$$
    \vskip-\lastskip\vskip -2pt}
\def\tablenote#1 #2\par{\begingroup \parindent=0.8em
    \abovedisplayshortskip=0pt\belowdisplayshortskip=0pt
    \noindent
    $$\hss\vbox{\hsize\tablewidth \hangindent=\parindent \hangafter=1 \noindent
    \hbox to \parindent{$^#1$\hss}\strut#2\strut\par}\hss$$
    \endgroup}

\def\L2{\ifmmode L_2\else $L_2$\fi}

\def\DeltaT{\ifmmode \Delta T\else $\Delta T$\fi}
\def\deltat{\ifmmode \Delta t\else $\Delta t$\fi}
\def\fknee{\ifmmode f_{\rm knee}\else $f_{\rm knee}$\fi}
\def\Fmax{\ifmmode F_{\rm max}\else $F_{\rm max}$\fi}
\def\solar{\ifmmode{\rm M}_{\mathord\odot}\else${\rm M}_{\mathord\odot}$\fi}
\def\Msolar{\ifmmode{\rm M}_{\mathord\odot}\else${\rm M}_{\mathord\odot}$\fi}
\def\Lsolar{\ifmmode{\rm L}_{\mathord\odot}\else${\rm L}_{\mathord\odot}$\fi}
\def\inv{\ifmmode^{-1}\else$^{-1}$\fi}
\def\mo{\ifmmode^{-1}\else$^{-1}$\fi}
\def\sup#1{\ifmmode ^{\rm #1}\else $^{\rm #1}$\fi}
\def\expo#1{\ifmmode \times 10^{#1}\else $\times 10^{#1}$\fi}
\def\,{\thinspace}
\def\lsim{\mathrel{\raise .4ex\hbox{\rlap{$<$}\lower 1.2ex\hbox{$\sim$}}}}
\def\gsim{\mathrel{\raise .4ex\hbox{\rlap{$>$}\lower 1.2ex\hbox{$\sim$}}}}

\def\simprop{\mathrel{\raise .4ex\hbox{\rlap{$\propto$}\lower 1.2ex\hbox{$\sim$}}}}
\def\deg{\ifmmode^\circ\else$^\circ$\fi}
\def\pdeg{\ifmmode $\setbox0=\hbox{$^{\circ}$}\rlap{\hskip.11\wd0 .}$^{\circ}
          \else \setbox0=\hbox{$^{\circ}$}\rlap{\hskip.11\wd0 .}$^{\circ}$\fi}
\def\arcs{\ifmmode {^{\scriptstyle\prime\prime}}
          \else $^{\scriptstyle\prime\prime}$\fi}
\def\arcm{\ifmmode {^{\scriptstyle\prime}}
          \else $^{\scriptstyle\prime}$\fi}
\newdimen\sa  \newdimen\sb
\def\parcs{\sa=.07em \sb=.03em
     \ifmmode \hbox{\rlap{.}}^{\scriptstyle\prime\kern -\sb\prime}\hbox{\kern -\sa}
     \else \rlap{.}$^{\scriptstyle\prime\kern -\sb\prime}$\kern -\sa\fi}
\def\parcm{\sa=.08em \sb=.03em
     \ifmmode \hbox{\rlap{.}\kern\sa}^{\scriptstyle\prime}\hbox{\kern-\sb}
     \else \rlap{.}\kern\sa$^{\scriptstyle\prime}$\kern-\sb\fi}
\def\ra[#1 #2 #3.#4]{#1\sup{h}#2\sup{m}#3\sup{s}\llap.#4}
\def\dec[#1 #2 #3.#4]{#1\deg#2\arcm#3\arcs\llap.#4}
\def\deco[#1 #2 #3]{#1\deg#2\arcm#3\arcs}
\def\rra[#1 #2]{#1\sup{h}#2\sup{m}}

\def\dots{\relax\ifmmode \ldots\else $\ldots$\fi}
\def\WHzsr{\ifmmode $W\,Hz\mo\,sr\mo$\else W\,Hz\mo\,sr\mo\fi}
\def\mHz{\ifmmode $\,mHz$\else \,mHz\fi}
\def\GHz{\ifmmode $\,GHz$\else \,GHz\fi}
\def\mKs{\ifmmode $\,mK\,s$^{1/2}\else \,mK\,s$^{1/2}$\fi}
\def\muKs{\ifmmode \,\mu$K\,s$^{1/2}\else \,$\mu$K\,s$^{1/2}$\fi}
\def\muKRJs{\ifmmode \,\mu$K$_{\rm RJ}$\,s$^{1/2}\else \,$\mu$K$_{\rm RJ}$\,s$^{1/2}$\fi}
\def\muKHz{\ifmmode \,\mu$K\,Hz$^{-1/2}\else \,$\mu$K\,Hz$^{-1/2}$\fi}
\def\MJysr{\ifmmode \,$MJy\,sr\mo$\else \,MJy\,sr\mo\fi}
\def\MJysrmK{\ifmmode \,$MJy\,sr\mo$\,mK$_{\rm CMB}\mo\else \,MJy\,sr\mo\,mK$_{\rm CMB}\mo$\fi}
\def\microns{\ifmmode \,\mu$m$\else \,$\mu$m\fi}

\def\muK{\ifmmode \,\mu$K$\else \,$\mu$\hbox{K}\fi}
\def\microK{\ifmmode \,\mu$K$\else \,$\mu$\hbox{K}\fi}
\def\muW{\ifmmode \,\mu$W$\else \,$\mu$\hbox{W}\fi}
\def\kms{\ifmmode $\,km\,s$^{-1}\else \,km\,s$^{-1}$\fi}
\def\kmsMpc{\ifmmode $\,\kms\,Mpc\mo$\else \,\kms\,Mpc\mo\fi}
\providecommand{\sorthelp}[1]{}

\pdfoutput=1

\begin{document}

\title{Joint power spectrum and voxel intensity distribution forecast on the CO luminosity function with COMAP}
\shorttitle{Joint PS and VID forecast for COMAP}
\shortauthors{H. T. Ihle et al.}

\correspondingauthor{H. T. Ihle}
\email{h.t.ihle@astro.uio.no}
\author{H. T. Ihle}
\affiliation{Institute of Theoretical Astrophysics, University of Oslo, P.O.Box 1029 Blindern, 0315 Oslo, Norway}

\author{D. Chung}
\affiliation{Kavli Institute for Particle Astrophysics and Cosmology and Physics Department, Stanford University, Stanford, CA 94305, USA}
\author{G. Stein}
\affiliation{Department of Astronomy and Astrophysics, University of Toronto, 50 St. George St., Toronto, ON, M5S 3H4, Canada}
\affiliation{Canadian Institute for Theoretical Astrophysics, University of Toronto, 60 St. George St., Toronto, ON M5S 3H8, Canada}

\author{M. Alvarez}
\affiliation{Berkeley Center for Cosmological Physics, University of California, Berkeley, CA 94720, USA}
\author{J. R. Bond}
\affiliation{Canadian Institute for Theoretical Astrophysics, University of Toronto, 60 St. George St., Toronto, ON M5S 3H8, Canada}
\author{P. C. Breysse}
\affiliation{Canadian Institute for Theoretical Astrophysics, University of Toronto, 60 St. George St., Toronto, ON M5S 3H8, Canada}
\author{K. A. Cleary}
\affiliation{California Institute of Technology, Pasadena, CA 91125, USA}
\author{H. K. Eriksen}
\affiliation{Institute of Theoretical Astrophysics, University of Oslo, P.O.Box 1029 Blindern, 0315 Oslo, Norway}
\author{M. K. Foss}
\affiliation{Institute of Theoretical Astrophysics, University of Oslo, P.O.Box 1029 Blindern, 0315 Oslo, Norway}
\author{J. O. Gundersen}
\affiliation{Department of Physics, University of Miami, 1320 Campo Sano Avenue, Coral Gables, FL 33146, USA}
\author{S. Harper}
\affiliation{Jodrell Bank Centre for Astrophysics, School of Physics and Astronomy, The University of Manchester, Oxford Road, Manchester, M13 9PL, U.K.}
\author{N. Murray}
\affiliation{Canadian Institute for Theoretical Astrophysics, University of Toronto, 60 St. George St., Toronto, ON M5S 3H8, Canada}
\author{H. Padmanabhan}
\affiliation{Institute for Particle Physics and Astrophysics, ETH Zurich, Wolfgang-Pauli-Strasse 27, CH 8093 Zurich, Switzerland}
\author{M. P. Viero}
\affiliation{Kavli Institute for Particle Astrophysics and Cosmology and Physics Department, Stanford University, Stanford, CA 94305, USA}
\author{I. K. Wehus}
\affiliation{Institute of Theoretical Astrophysics, University of Oslo, P.O.Box 1029 Blindern, 0315 Oslo, Norway}

\collaboration{(COMAP collaboration)}

\begin{abstract} 

 We develop a framework for joint constraints on the CO
  luminosity function based on power spectra (PS) and voxel intensity
  distributions (VID), and apply this to simulations of COMAP, a CO intensity 
  mapping experiment. This Bayesian framework is based on a 
  Markov chain Monte Carlo (MCMC) sampler coupled to a Gaussian likelihood with a joint
  PS + VID covariance matrix computed from a large number of fiducial
  simulations, and re-calibrated with a small number of simulations per
  MCMC step. The simulations are based on dark matter halos from fast
  peak patch simulations combined with the
  $L_\text{CO}(M_\text{halo})$ model of \cite{Li}. We find that the
  relative power to constrain the CO luminosity function depends on
  the luminosity range of interest. 
   In particular, the VID is more sensitive at large 
  luminosities, while the PS and the VID are both competitive at 
  small and intermediate luminosities.
  The joint analysis is superior to using either
  observable separately. When averaging over CO
  luminosities ranging between $L_\text{CO} = 10^4-10^7L_\odot$, and over 10 cosmological 
  realizations of COMAP Phase 2, the uncertainties (in dex) are larger by 58\% and
  30\% for the PS and VID, respectively, when compared to the joint analysis (PS + VID). This method is
  generally applicable to any other random field, with a complicated likelihood, as long a fast
  simulation procedure is available.
  
\end{abstract}


\section{Introduction}

Intensity mapping \citep{madau1997, Battye2004, Peterson2006, Loeb2008} appears
promising for mapping large 3D volumes cheaply in a
relatively short period of time, using specific bright emission lines as matter
tracers. This is an interesting avenue for advancing precision cosmology, with a 
multitude of ongoing efforts \citep{LIM2017}, 
following on the successes of the CMB field in the last few decades. 
One such line intensity mapping experiment currently under construction is called
the CO Mapping Array Pathfinder (COMAP; \citealp{Kieran,Li}), which aims to
observe frequencies between 26 and 34~GHz, corresponding to redshifted
CO line emission from the epoch of galaxy  assembly (redshifts between
$z=2.4$ and 3.4) for the CO $J$=1$\rightarrow$0 line at 115~GHz rest
frequency, and CO emission from the epoch of reionization
($z=5.8$--6.7) for the CO $J$=2$\rightarrow$1 line at 230~GHz rest
frequency. 

One important scientific target for studying and understanding the
epoch of galaxy  assembly, the main goal of the first COMAP phase, is
the so-called CO luminosity function, which measures the number  density of CO
emitters as a function of luminosity. Several methods for extracting
this function from real data 
have already been suggested in the literature, and
the most prominent of these is the power spectrum (PS) approach, for
instance as implemented by \cite{Li}. A second complementary method is
the 1-point function, or voxel intensity distribution (VID),
$\mathcal{P}(T)$, as suggested by \cite{Breysse1, Breysse2}.

In this paper, we consider the prospect of
combining the VID and PS approaches when constraining the CO
luminosity function, and we study this approach within the context of
the COMAP experiment. To do so, we first define a joint likelihood
that includes both the VID and the PS, and construct a joint
covariance matrix for both observables. This covariance matrix is
constructed from a large set of dark matter (DM) light-cone halo
catalogs from so-called ``peak patch'' cosmological simulations \cite{bond1996, Stein2018},
coupled to an empirical $L_{\mathrm{CO}}(M_\text{halo})$ model
\citep{Li} that infers CO luminosities, $L_{\mathrm{CO}}$, from DM
halo masses, $M_\text{halo}$. We then investigate the posterior
distribution of the resulting model parameters for each of the first
two anticipated phases of the COMAP experiment (see Table \ref{tab:comap}). Finally, we compare the
constraints on the CO luminosity function derived from joint
PS and VID measurements to those
obtained from the PS or VID separately.

\section{Idealized simulations of the COMAP experiment}

\begin{table}[tb]
\begingroup
\newdimen\tblskip \tblskip=5pt
\caption{Experiment setup for the two COMAP phases.\label{tab:comap}}
\vskip -4mm
\footnotesize
\setbox\tablebox=\vbox{
\newdimen\digitwidth
\setbox0=\hbox{\rm 0}
\digitwidth=\wd0
\catcode`*=\active
\def*{\kern\digitwidth}
\newdimen\signwidth
\setbox0=\hbox{+}
\signwidth=\wd0
\catcode`!=\active
\def!{\kern\signwidth}
\newdimen\decimalwidth
\setbox0=\hbox{.}
\decimalwidth=\wd0
\catcode`@=\active
\def@{\kern\signwidth}
\halign{\hbox to 1.8in{#\leaderfil}\tabskip=2.0em&
    \hfil#\hfil\tabskip=1em&
    \hfil#\hfil\tabskip=1em\cr
\noalign{\vskip 3pt}
\omit\hfil\sc Parameter\hfil& COMAP1 & COMAP2 \cr 
\noalign{\vskip 5pt\hrule\vskip 5pt}
System temperature, $T_\text{sys}$ [K] & 40 & 40\cr
Number of feeds & 19 & 95\cr
Beam FWHM (arcmin) & 4 & 4 \cr
Frequency band [GHz] & 26--34 & 26--34 \cr
Channel width, $\delta\nu$ (MHz) & 15.6 & 15.6 \cr
Observing time [h] & 6000 & 9000 \cr
Noise per voxel [$\mu$K] & 11.0 & 8.0 \cr
Field size [deg${}^2$] & 2.25 &  2.25\cr
Number of fields & 1 & 4\cr
}}
\endPlancktablewide 
\endgroup
\end{table}

We start our discussion by reviewing some central properties of the
COMAP experiment, focusing in particular on those required for
generating representative yet computationally affordable
simulations. For convenience, these properties are summarized in
Table~\ref{tab:comap}.

In Phase 1 COMAP will employ a single telescope equipped with 19
single-polarization detectors, each with 512 frequency channels with
width $\delta\nu \approx$ 15.6~MHz\footnote{Higher spectral resolutions are available, but
  these are most likely useful only for systematics mitigation rather
  than science, due to limited signal-to-noise per voxel.} covering frequencies between 26 GHz
and 34~GHz. The system
temperature is expected to be around $T_\text{sys} \approx 40$ K, and
the angular resolution corresponding to a Gaussian beam with $4'$ full width 
at half maximum (FWHM). We anticipate two years of
observation time targeting a single field of 1.5 deg $\times$ 1.5 deg
close to the north celestial pole, and we assume a conservative
observing efficiency of 35\%, for a total of 6000 hours of total integration time on the field.

In Phase 2, the experiment will be expanded to five telescopes with
the same setup as in Phase 1, and observe for three additional
years. In this phase, we assume that the observation time will be
split between four fields of the same size as in Phase 1. The two
COMAP phases will be referred to as COMAP1 and COMAP2 in the
following.

\subsection{Noise}
The simulations used in this work consist of two components only,
namely the target CO signal and random white noise with properties
corresponding to the parameters described above. Explicitly, the noise
per voxel, is given by 
\begin{equation}
 \sigma_T = \frac{T_\text{sys}}{\sqrt{\tau\,\delta \nu}} = \frac{T_\text{sys}\sqrt{N_\text{pixels}}}{\sqrt{\tau_\text{tot} e_\text{obs} N_\text{feeds} \delta \nu}}, 
\end{equation}
where $T_\text{sys}$ is the system temperature, $\tau$ is the observation time per pixel, 
$\tau_\text{tot}$ is the total observation time
, $e_\text{obs}$ is the observation efficiency, $N_\text{feeds}$ is the 
number of feeds, $N_\text{pixels}$ is the number of pixels, and 
$\delta \nu$ is the frequency resolution. This gives us 
$\sigma_T \approx 11\, \mu \text{K and } 8\, \mu \text{K}$, 
for the COMAP1 and COMAP2 phases respectively. 
For simplicity we assume that the noise is evenly distributed over all voxels.

A voxel is the 3D equivalent of a pixel. Two of the dimensions correspond to 
a regular pixel on the sky, while the third dimension corresponds to a 
small range of redshifts from where line emmision would redshift into a 
given frequency bin of our instrument. 

Both instrumental systematics and astrophysical
foreground contamination are neglected in the following. However,
since our estimator is inherently simulation-based, these effects can
be added at a later stage, when a sufficiently realistic instrument
model is available. For discussion of foreground contamination in similar
line intensity surveys see e.g. \cite{daCunha2013, BreysseMask,Breysse1, Chung2017}.

\subsection{Dark matter simulations}
The signal component is based on the peak patch DM halo approach
described by \cite{bond1996, Stein2018}, coupled to the $L_{\mathrm{CO}}(M_\text{halo})$ model
presented by \cite{Li}. Additionally, we adopt the same
cosmological parameters as the \cite{Li} analysis for the dark matter simulations: $\Omega_m =
0.286$, $\Omega_\Lambda = 0.714$, $\Omega_b = 0.047$, $h = 0.7$,
$\sigma_8 = 0.82$, and $n_s$ = 0.96.

The DM simulations in this paper were created using the peak patch method of \cite{bond1996, Stein2018}. 
To cover the full redshift range of the COMAP experiment we simulated a volume of (1140 Mpc)$^3$ using 
a particle-mesh resolution of $N_\textrm{cells}=4096^3$. Projecting this onto the sky results in a 
$9.6^{\circ} \times 9.6^{\circ}$ field of view covering the redshift range $2.4 < z < 3.4$, 
with a minimum dark matter halo mass of 
2.5$\times$10$^{10}M_\odot$. 

The resulting halo catalog contains roughly 54 million 
halos, each with a position, a velocity, and a mass. The peak patch method can simulate 
continuous light-cones on-the-fly, so stitching snapshots together was not required to 
create the light-cone. Although peak patch simulations result in quite accurate halo 
masses, the dark matter halo catalogs were additionally mass corrected by abundance 
matching along the lightcone to \cite{Tinker} to ensure statistically the same mass function as the 
simulations used in the \cite{Li} analysis.  For a detailed study
of the clustering properties of peak patch simulations and other approximate 
methods see \cite{mock1, mock2, mock3}.

A single run required 900 seconds of 
compute time on 2048 Intel Xeon EE540 2.53 GHz CPU cores of the Scinet-GPC cluster, 
with a memory footprint of $\simeq$~2.4 TB. This efficiency of the peak 
patch method allowed for 161 independent realizations of the full 1140Mpc, 
$N_\textrm{cells}=4096^3$ volume, taking a total compute time of only $\sim$~82,000 CPU hours,
over 3 orders of magnitude faster when compared to an N-body method of equivalent size. 

\subsection{Converting to CO brightness temperature}
There are many approaches in the literature for estimating the expected CO signal based on DM halos  
\citep[e.g.][]{Righi2008,Obreschkow2009,Visbal2010,Lidz2011,Carilli2011,Gong2011,Fu2012,Pullen2013,Carilli2013,Breysse2014,Greve2014, Mashian2015,Li,Padmanabhan2017}, 
with resulting estimates of the CO-luminosities spanning roughly an order of magnitude.

Here we adopt the model described by
\cite{Li}, to convert from simulated light-cones populated with DM halos to
observed CO brightness temperature. This model is defined by a set of parametric relations
between DM halo masses, star formation rates (SFR), infrared (IR)
luminosities, $L_\text{IR}$, and CO-luminosities, $L_\text{CO}$. 

The model uses the results from \cite{Behroozi13a,Behroozi2013} to obtain average SFR
from DM halo masses, and adds an additional log-normal scatter on top of 
the average, determined by $\sigma_{\mathrm{SFR}}$.  IR luminosities 
are then obtained through the relation
\begin{equation}
 \mathrm{SFR} = \delta_{\mathrm{MF}} \times 10^{-10} L_{\mathrm{IR}}. 
\end{equation}
Further, to obtain CO luminosities, the relation
\begin{equation}
 \log L_{\mathrm{IR}} = \alpha \log L^\prime_{\mathrm{CO}} + \beta,
\end{equation}
is used before a second round of log-normal scatter is added, determined by
the parameter $\sigma_{L_{\mathrm{CO}}}$.

\begin{figure}
\centering
	\includegraphics[width=0.5\textwidth]{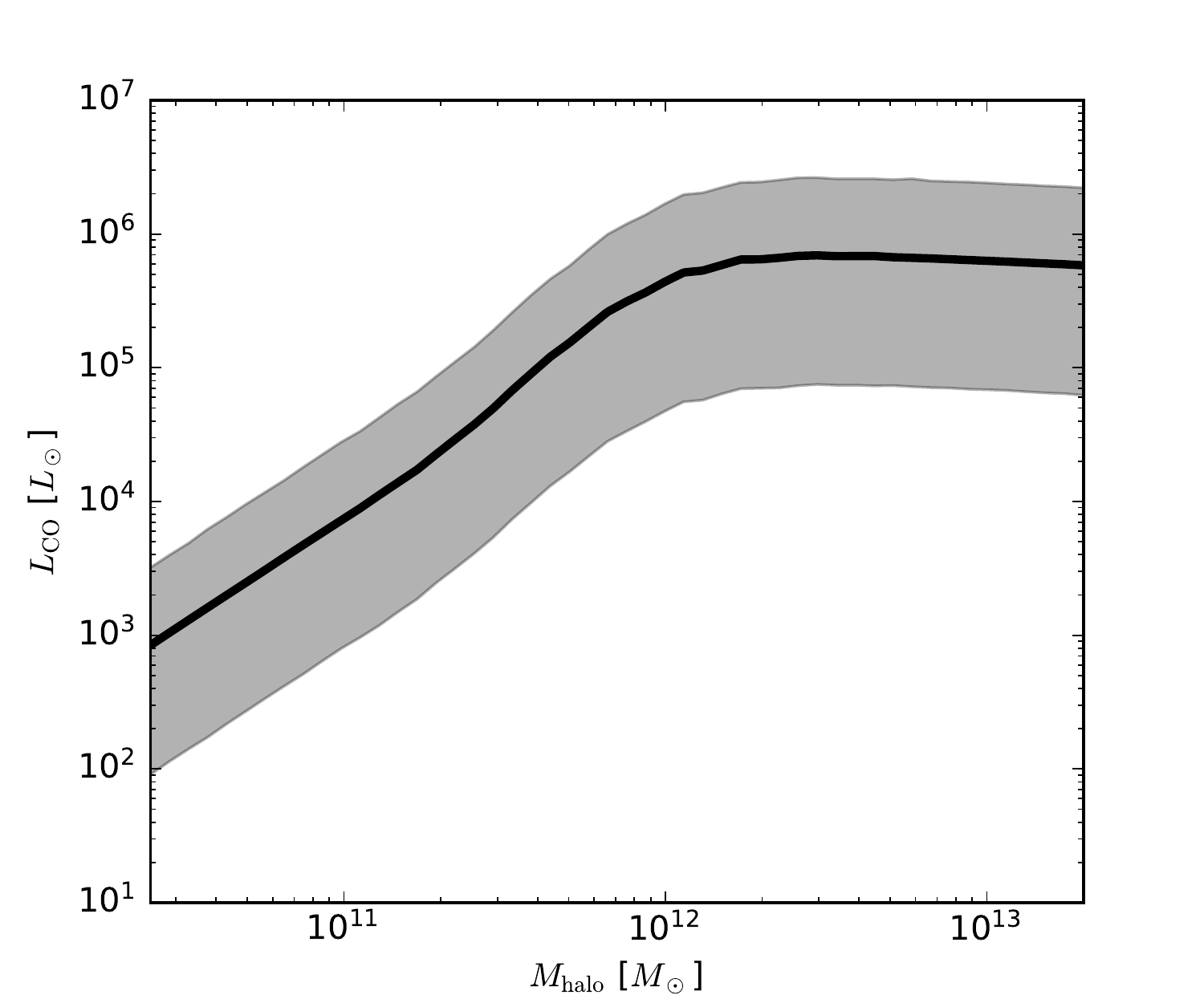}
  \caption{Plot of CO luminosity, $L_{\mathrm{CO}}$, 
  as a function of dark matter halo mass, $M_\text{halo}$, in the \cite{Li} model.
  Here $(\sigma_\mathrm{SFR}, \log
\delta_\text{MF}, \alpha, \beta, \sigma_{L_{\mathrm{CO}}}) = (0.3, 0.0, 1.17, 0.21, 0.3)$
(our fiducial model), and we have evaluated the function at redshift $2.9$. The solid line 
corresponds to the mean relation with no scatter added, while the shaded region
corresponds to the $95\%$ confidence intervals after adding log-normal scatter at the two 
appropriate steps. 
  \label{fig:L_Mhalo}}
\end{figure}

This gives us a $L_\mathrm{CO}(M_\mathrm{halo})$ model with five free parameters, 
$\theta = \{\sigma_\mathrm{SFR}, \log
\delta_\text{MF}, \alpha, \beta, \sigma_{L_{\mathrm{CO}}}\}$.  The relation
between $L_{\mathrm{CO}}$ and $M_\text{halo}$, for our fiducial model parameters, is shown in 
Figure \ref{fig:L_Mhalo}. 
For more discussion of the physical and observational motivation for this model, see the original 
paper \cite{Li}. 

This model is applied to each DM halo separately, and from the
resulting CO luminosities we create high-resolution maps by converting
the total luminosity in a given voxel into brightness
temperature. These maps were created using the publicly available 
\texttt{limlam\_mocker} code\footnote{https://github.com/georgestein/limlam\_mocker}. 
Next, we convolve these maps with the (Gaussian)
instrumental beam profile, degrade to the low-resolution voxel size
used in the analysis, and, finally, we add Gaussian uncorrelated
noise with standard deviation $\sigma_T$ as specified above. 

\section{Algorithms}

The ultimate goal of this work is to constrain cosmological and
astrophysical parameters from CO line intensity observations. The
computational engine for this work is a standard Metropolis 
Markov chain Monte Carlo (MCMC) sampler \citep[see e.g.,][]{gilks1995markov},
coupled to a posterior distribution
with a corresponding likelihood and prior. For this task to be
computationally tractable, though, the full CO line intensity data set
must first be compressed to a smaller set of observables that may be
modeled in terms of the desired astrophysical parameters, fully
analogous to how CMB sky maps are compressed to a CMB power spectrum
from which cosmological parameters are derived \citep[e.g.,][]{bond2000}. 
As described above, we
adopt the power spectrum and the voxel intensity distribution as
representative observables, each of which may be approximated in terms
of multivariate Gaussian random variables. However, in order to
perform a joint analysis of these two observables, we need to
construct their joint covariance matrix, and that is the primary goal
of this section. Before doing that, however, we review for
completeness each observable individually, and our posterior sampler
of choice, referring to relevant literature for full details.

\subsection{The power spectrum}
\label{sec:obs}

The estimated power spectrum, $P(k_i)$, is calculated simply by taking
the 3D Fourier transform of the temperature cube, binning the absolute
squared values of the Fourier coefficients according to the magnitude
of corresponding wave number $k$, and averaging over all the contributions
within each bin. For a Gaussian map, the Fourier components within
each bin follow a perfect normal distribution with mean zero and
variance given by the value of the power spectrum. For a non-Gaussian
field the distribution of the Fourier components is more complicated,
and thus the power spectrum does not contain all the statistical
information in the map. We expect the CO signal to form a highly 
non-Gaussian map, therefore, in this paper, we simply consider the power
spectrum as a useful observable that carries some, but far from all, of
the statistical information in the map.

As an observable, the power spectrum needs to be accompanied by a
covariance matrix $\xi^P_{ij} \equiv \text{Cov}(P(k_i), P(k_j))$ in
the analysis, since there are correlations between the power spectrum
at different $k$-values. 

\subsection{The voxel intensity distribution}
\label{sec:vid}
We consider the VID as another observable, complementary to
the PS, and more closely related to the luminosity function. 

We do not, unlike in many other works on $\mathcal{P}(D)$ analysis
(e.g. \cite{Lee2009, Glenn2010, Vernstrom2014, Breysse1, Leicht2018}), try to estimate the 
VID analytically, rather we estimate it based on simulations. This allows 
us to fully take into account the effects of the beam, clustering and covariance
between temperature bins in a very straightforward manner.

We consider two contributions to the VID, namely the CO signal itself
and the instrumental noise. Together they result in the the full probability 
distribution of voxel temperatures, $\mathcal{P}(T)$, where $T$
is the observed brightness temperature from a voxel. Since
we, in this paper, assume the noise to be uniformly distributed over all
voxels in the observed field, and the CO signal itself is
statistically homogeneous and isotropic, the total probability
distribution, $\mathcal{P}(T)$, is
the same across all voxels.

The basic observable related to the VID are the temperature bin counts (i.e. the histogram of voxel
temperatures), $B_i$. The expectation value of these are given by the
VID itself,
\begin{equation}
 \langle B_i \rangle = N_\text{vox} \int_{T_i}^{T_{i+1}} \mathcal{P}(T) dT, 
\end{equation}
where $N_\text{vox}$ is the number of voxels
observed and $B_i$ is the number of voxels with a temperature between $T_i$
and $T_{i+1}$.

If the temperatures of all the voxels that we sample were completely
independent, then each of the voxel bins would be approximately
independent, and follow a binomial distribution with variance
$\text{Var}_\text{ind}(B_i) = \langle B_i \rangle(1 - \langle B_i
\rangle/N_\text{vox})$. However, even in this ideal case they would not be perfectly
independent. We only have a finite number of voxels, and,
therefore, if one bin contains a number of voxels above average, then
the other bins must have a number of bins lower than the average.

In general, the samples will not be independent for many reasons,
including correlated sky or noise structures and processing effects,
and we therefore need the full covariance matrix between bins,
$\xi^B_{ij} \equiv \text{Cov}(B_i, B_j)$. This covariance matrix will
depend on 
the DM density field, the CO bias, and the luminosity
function, and we will estimate it using simulations.

\subsection{The joint PS+VID covariance matrix}
\label{sec:covmat}

The main missing component in the above method is definition of a
joint power spectrum and voxel intensity distribution covariance matrix. 
By having access to the
computationally cheap yet realistic Monte Carlo simulations described
above, we can
approximate this matrix by simulations.
In addition to giving us covariance matrices to do our analysis, this
also allows us to check under what conditions the full covariance
matrix is necessary, and when we can get away with assuming that
individual samples are independent.

In this paper, we start out with 161 independent simulated light-cone
cubes of dark matter halos, each covering about 
$9.6^{\circ} \times 9.6^{\circ}$ on the
sky, and a frequency range between 26 and 34~GHz, corresponding to
redshifts between 2.4 and 3.4. The frequency dimension is divided
equally into 512 frequency bins, each spanning
$\delta\nu\approx15.6~\mathrm{MHz}$, corresponding to a redshift
resolution of $\delta z\approx0.002$. Since the COMAP field only spans
$1.5^{\circ} \times 1.5^{\circ}$ on the sky, we sub-divide each of the 
$9.6^\circ\times9.6^\circ$ light-cone cubes, after beam convolution,
into 36 square fields
each covering $1.5^{\circ} \times
1.5^{\circ}$
, resulting in a total of 5796 semi-independent sky
realizations. The final pixelisation of these maps is a $22\times22$ 
grid of square pixels, resulting in a pixel size of
$\delta \theta
\approx 4.1'$. To these maps, we add uniformly distributed white noise at
the appropriate levels for the COMAP1 and COMAP2 experiment setups
described above.

When choosing the pixel size to use for the analysis, we follow 
\cite{Vernstrom2014}. They show that, for $\mathcal{P}(D)$ analysis, choosing a pixel 
size to be equal to the FWHM of the beam is a good trade-off between picking
a small pixel size to include the maximal information, and choosing a larger 
pixel size to reduce the pixel to pixel correlations induced by the beam. 

We combine our two observables into a joint one-dimensional vector of
the form
\begin{equation}
 d_i = (P_{k_i}, B_i), 
\end{equation}
where $P_{k_i}$ is the binned power spectrum, and $B_i$ are the temperature bin counts. 
Let us first consider the ideal case in which all elements in this
vector are independent, and the Fourier components are approximately Gaussian. 
In that case we can compute the expected variance, which we will simply 
call the \emph{independent variance}, analytically,
\begin{align}
 \text{Var}_\text{ind}(P_{k_i}) &= \langle P_{k_i}\rangle^2 / N_\text{modes}(k_i), \\
 \text{Var}_\text{ind}(B_i) &= \langle B_i \rangle(1 - \langle B_i \rangle/N_\text{vox}) \approx \langle B_i \rangle,
\end{align}
where $N_\text{modes}(k_i)$ is the number of modes in the $i^\text{th}$ $k-$bin 
and where we have introduced the notation $\text{Var}_\text{ind}(d_i)$ for
this conditionally independent variance.

With this notation in hand, we define a ``pseudo-correlation matrix''
as
\begin{equation}
 c_{ij} \equiv \frac{\xi_{ij}}{\sqrt{\text{Var}_\text{ind}(d_i)\text{Var}_\text{ind}(d_j)}},
\end{equation}
where, as in Sect.~\ref{sec:mcmc}, $\xi_{ij}$ is the full covariance
matrix. Note that $c_{ij}$ is the exact correlation matrix in the
limit that $\text{Var}_\text{ind}(d_i)$ is the true full variance. An
important advantage of the pseudo-correlation matrix, however, is the
fact that $\text{Var}_\text{ind}(d_i)$ may be estimated directly from
the average data itself, and this is required for our MCMC procedure
to be sufficiently fast.

The full covariance matrix $\xi$ is estimated for the model
described by \cite{Li}, adopting the fiducial parameters
$\theta_0$, using the set of 5796 simulations described above. However,
for the MCMC sampler described in Sect.~\ref{sec:mcmc}, we actually
need the full covariance matrix, corresponding to different model 
parameters $\theta$, at each step in the Markov chain. 
Generating the full covariance matrix
with the above procedure at each MC step is clearly not
computationally feasible, and we therefore need to approximate this
somehow.

With regard to this last point, we introduce the following proposal: 
we assume that the full covariance matrix scales, under a change of 
model parameters from $\theta_0$ to $\theta$, the same way as the 
independent variance, $\text{Var}_\text{ind}(d_i)$, does
\begin{equation}\label{eq:xi_scaling}
 \hat{\xi}_{ij}(\theta) \approx \xi_{ij}(\theta_0)
 \frac{\sqrt{\text{Var}_\text{ind}^{\theta}(d_i)\text{Var}_\text{ind}^{\theta}(d_j)}}{\sqrt{\text{Var}_\text{ind}^{\theta_0}(d_i)\text{Var}_\text{ind}^{\theta_0}(d_j)}},
\end{equation}
where $\text{Var}_\text{ind}^{\theta_0}(d_i)$ is the
independent variance for the fiducial model, and
$\text{Var}_\text{ind}^{\theta}(d_i)$ is the independent
variance for arbitrary parameters $\theta$. Since this latter function
only depends on the average quantities $\langle d_i \rangle$, it is
computationally straightforward to compute $\hat{\xi}_{ij}(\theta)$ at
any position in a MCMC sampler. Note also that
$\hat{\xi}_{ij}(\theta)$ is, by construction, positive definite, as
required for a proper covariance matrix.

For a noise dominated experiment, where all samples are 
approximately independent, the independent variance, 
$\text{Var}_\text{ind}(d_i)$, is the correct variance, and 
Eq.~\ref{eq:xi_scaling} is the correct scaling of the 
covariance matrix. However, we use this scaling as a first 
approximation even in cases where there is some covariance in the 
data. 

Intuitively, Eq.~\ref{eq:xi_scaling} is equivalent to postulating that
the pseudo-correlation matrix, $c_{ij}$, is approximately constant (i.e. independent of the 
specific parameters in question). 
For real-world applications, we recommend testing this assumption
explicitly by computing the covariance matrix by brute force
simulation for a few extreme parameter combinations drawn from the
Markov chains produced during the analysis.

The above prescription applies straightforwardly to single-field
observations as, for instance, planned for COMAP1. In contrast, COMAP2
will, under our assumptions, span $N=4$ independent but statistically
identical fields. Since the mean vector of observables evaluated
across those four fields equals the average of the four corresponding
independent observable vectors, the full covariance matrix is simply
given by the single field covariance matrix divided by the number of fields:
\begin{equation}
 \xi_{ij}^{N\, \mathrm{field}} = \frac{\xi_{ij}^{\text{1 field}}}{N}.
\end{equation}
Note that $c_{ij}$ then, assuming the fields are of the same size,
only depends on the noise level per field, 
so for a given noise level per field, $c_{ij}$ is independent of 
the number of fields.

Finally, we note that the total number of degrees of freedom in our
joint PS+VID statistic is in this paper equal to 45, corresponding to
20 power spectrum bins and 25 VID bins. For this number of degrees of 
freedom, a set of 5796 (semi-independent) simulations provides a very good estimate of 
all numerically
dominant components of the covariance matrix, including both the
diagonal and the leading off-diagonal modes, and $\xi_{ij}$ is well
conditioned.

\subsection{Posterior mapping by MCMC}
\label{sec:mcmc}
As previously mentioned, we use 
a MCMC algorithm to sample
from the posterior distribution of the $L_{\mathrm{CO}}(M_\text{halo})$ model
parameters, $\theta = \{\sigma_\mathrm{SFR}, \log \delta_\text{MF}, \alpha, \beta, \sigma_{L_{\mathrm{CO}}}\}$. This posterior distribution is, as usual, given by
Bayes' theorem,
\begin{equation}
P(\theta|d) \propto P(d_i|\theta) P(\theta),
\end{equation}
where $d$ represents our compressed data set, $P(d|\theta)$ is the
likelihood defined below, and $P(\theta)$ is some set of priors. 
We use the \emph{emcee} package
\citep{emcee}
and its implementation of an affine-invariant ensemble MCMC algorithm, with 142 walkers. 

We use a burn-in period of 1000 steps, and use the next 1000 steps 
for the posterior estimation. 

We assume a Gaussian likelihood for our observables $d_i$
of the form (up to an additive constant)

\begin{equation}\label{eq:likelihood}
 -2\ln P(d|\theta) = \sum_{ij}[d_i -\langle d_i \rangle] (\xi^{-1})_{ij}[d_j - \langle d_j \rangle] + \ln|\xi|,
\end{equation}

where the means $\langle d_i \rangle$ depend on the model parameters
$\theta$, and the covariance matrix $\xi_{ij}$ is approximated by the
expression given in Eq.
\ref{eq:xi_scaling}. (Note that we do not need to assume that the
low-level data are Gaussian, but only that the compressed observables
may be well approximated by a multi-variate Gaussian distribution. Due
to the central limit theorem, this is in practice very often an
excellent approximation.)

For both the power spectrum and the low and intermediate temperature
VID bins, for which there is a large number of voxel counts per bin,
this Gaussian approximation holds to a high degree. However, for the
highest VID temperature bins, where there are only a few voxel counts
per bin, the discrete nature of the bin count may become relevant, and
the full binomial distribution should in principle be taken into
account. 
This effect can however also be remedied easily by increasing
the bin width, albeit at the cost of slight loss of information, as is 
suggested in \cite{Vernstrom2014}, and
we therefore neglect it in the following, as our primary focus is the
dominant Gaussian component of the likelihood. 
A more thorough
analysis may take this issue into account either analytically or by
simulations. 

\begin{figure*}[t]
  \center
  \includegraphics[width=0.48\textwidth]{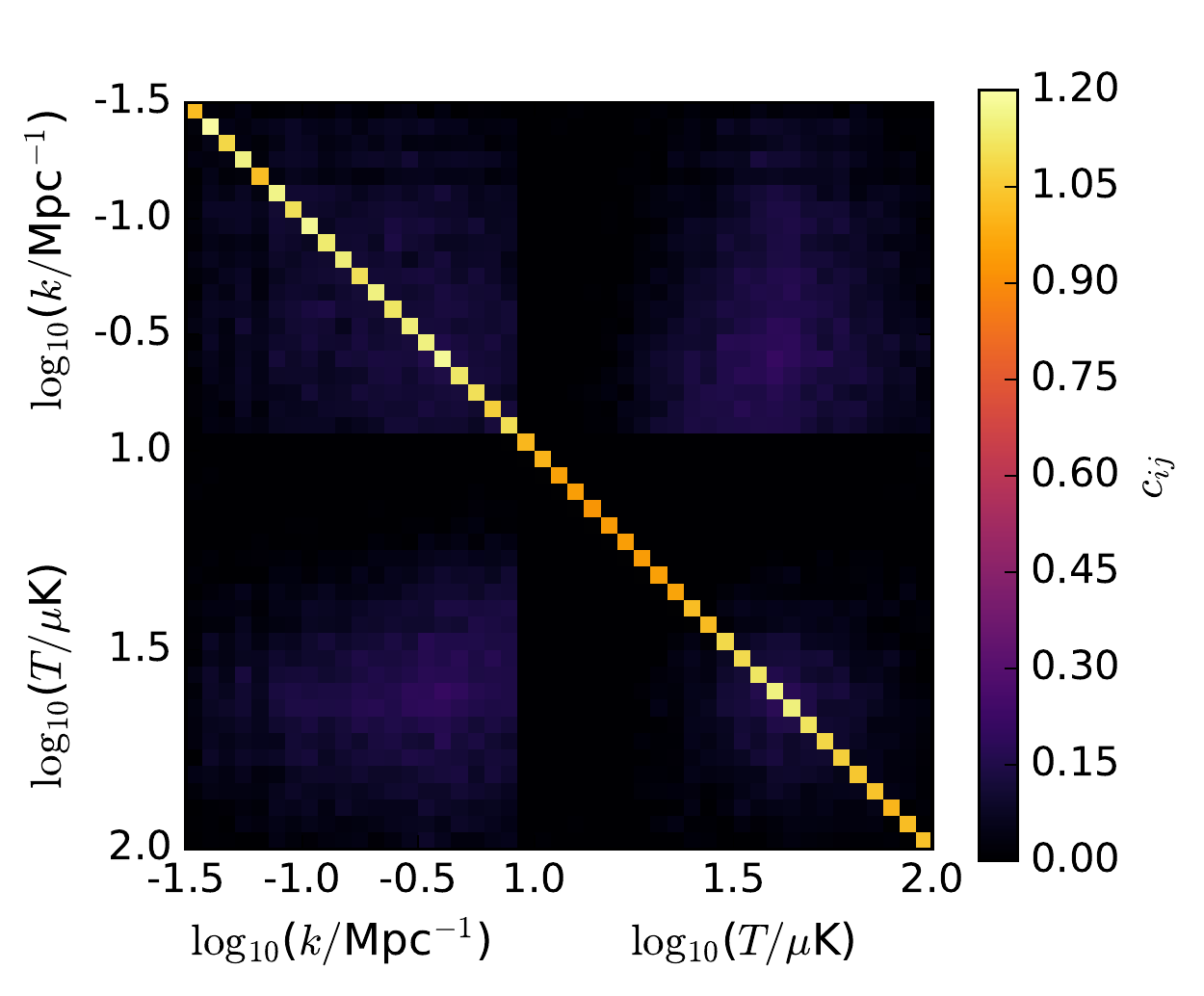}
  \includegraphics[width=0.48\textwidth]{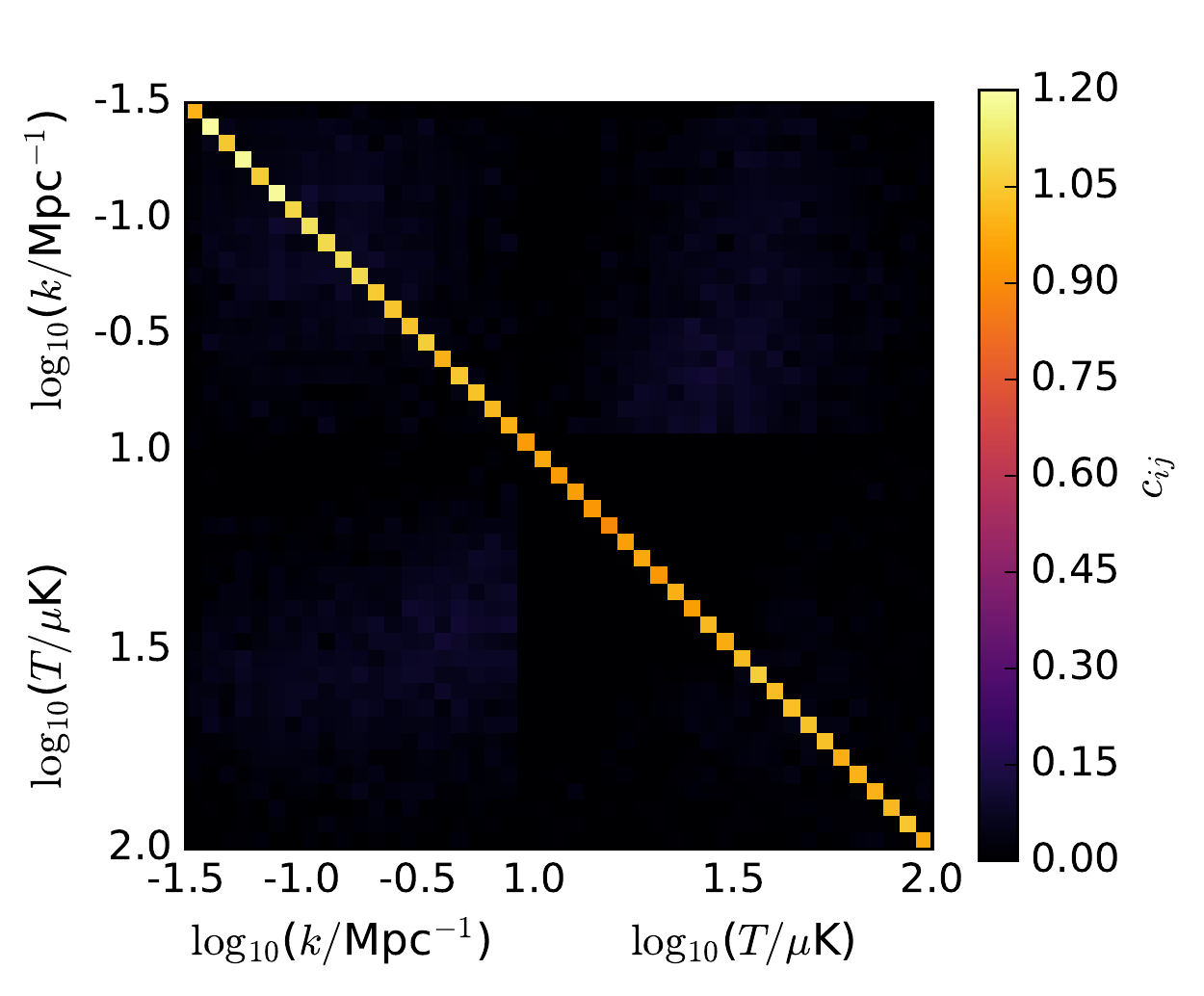}\\
  \includegraphics[width=0.48\textwidth]{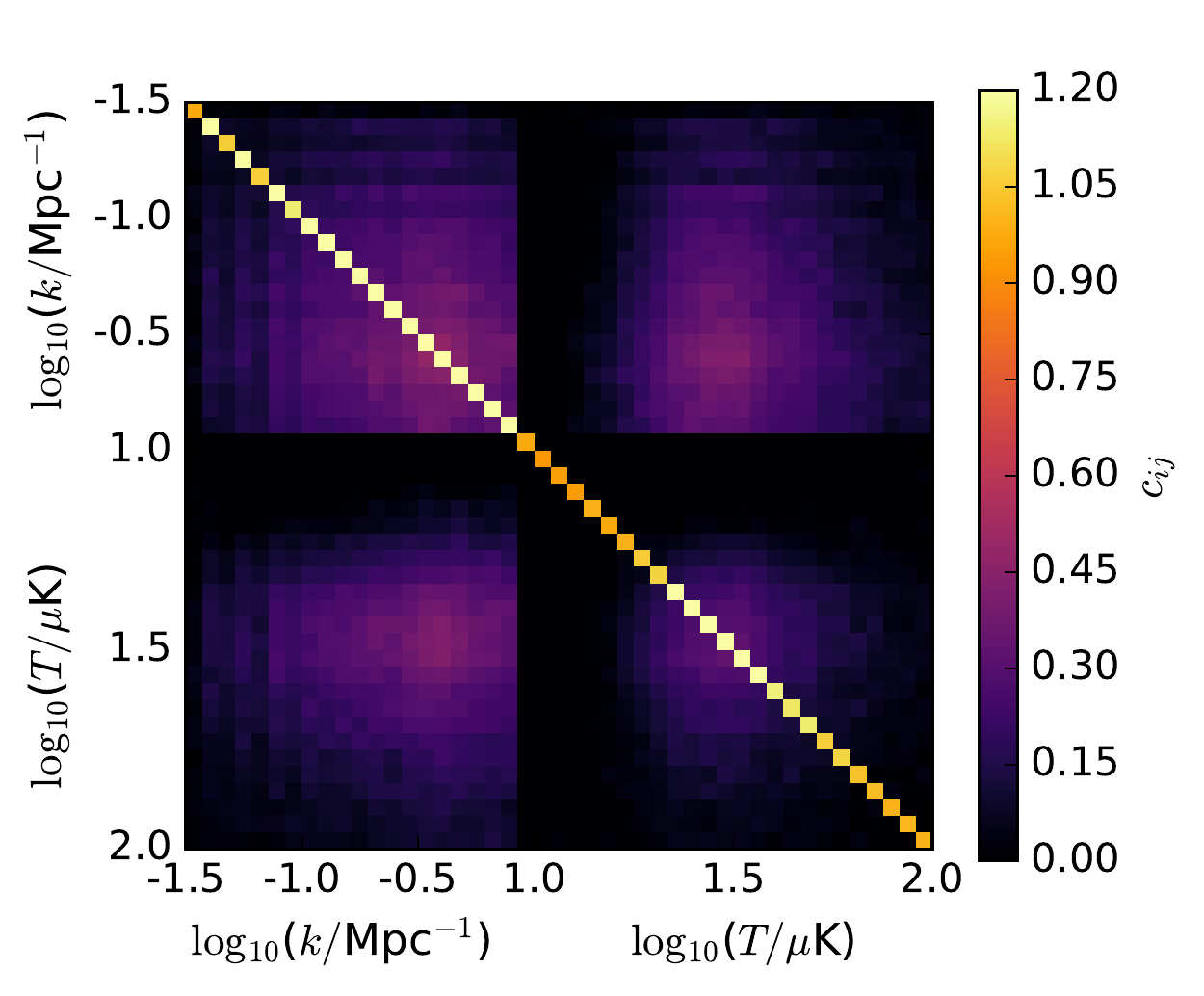}
  \includegraphics[width=0.48\textwidth]{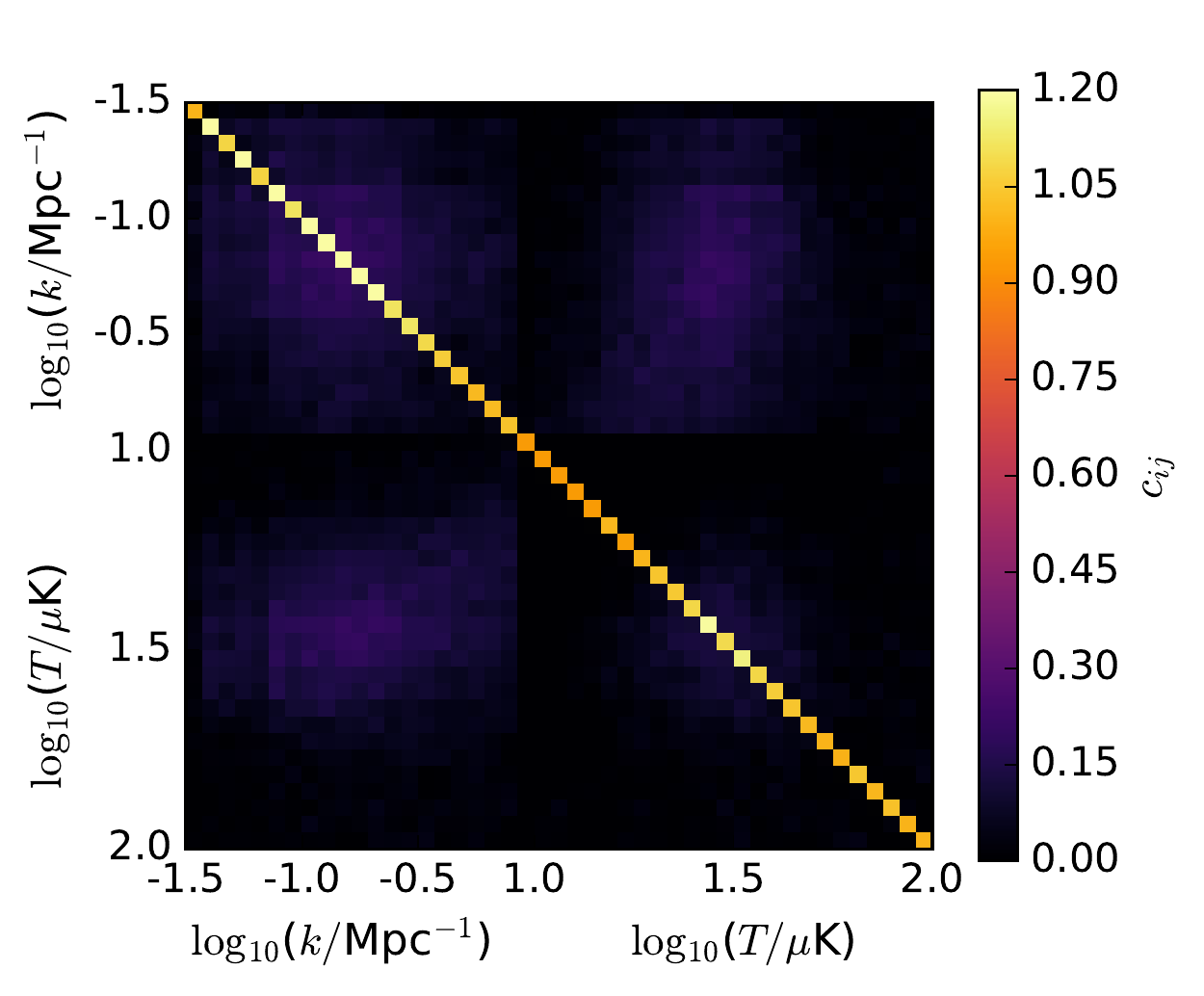}\\
  \includegraphics[width=0.48\textwidth]{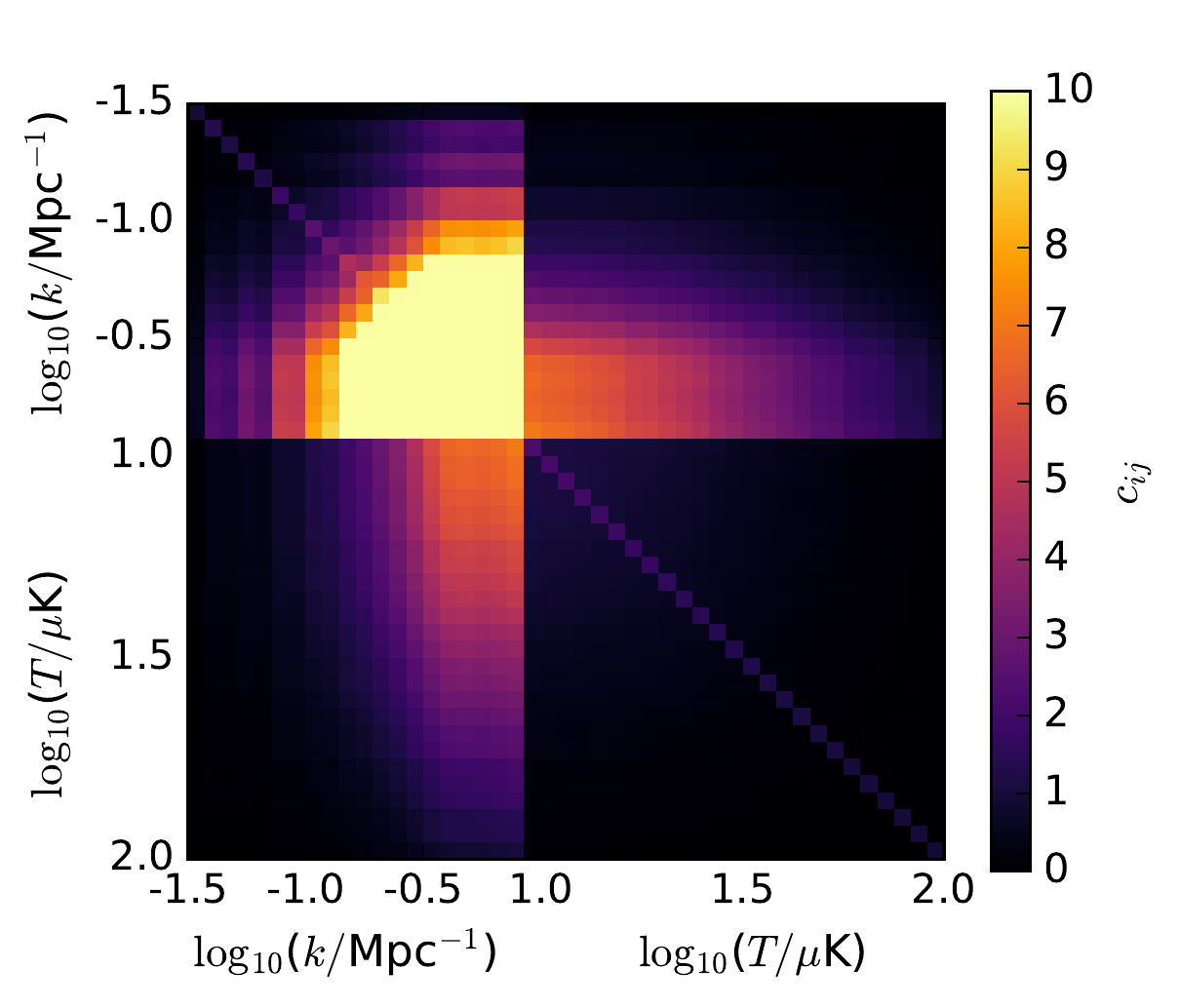}
  \includegraphics[width=0.48\textwidth]{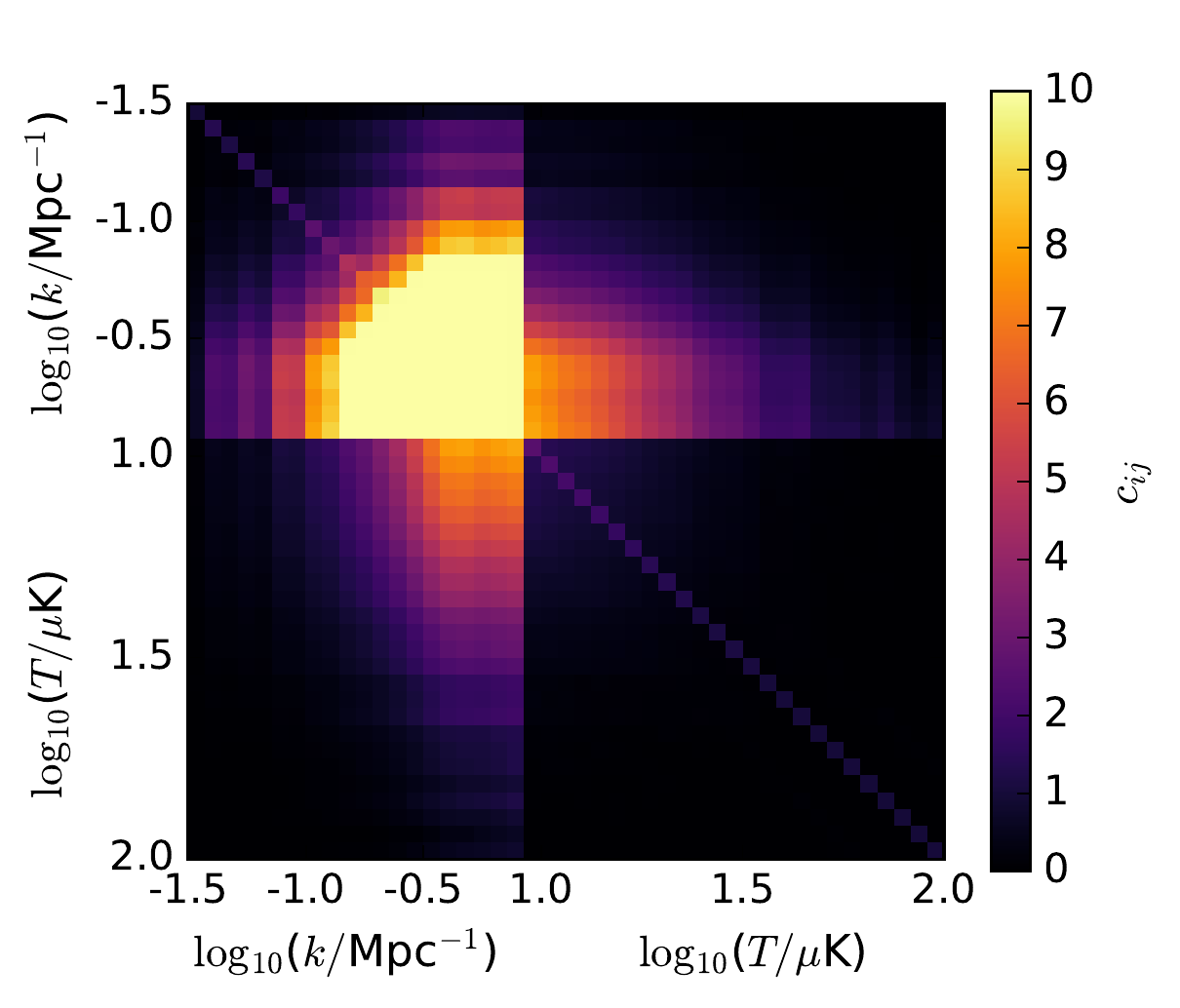}
  \caption{Estimated pseudo-correlation matrix of observables
    $d_i$, $c_{ij} = \text{Cov}(d_i,
    d_j)/(\sqrt{\text{Var}_\text{ind}(d_i)\text{Var}_\text{ind}(d_j)})$,
    based on simulated maps with and without noise. The first block
    in each matrix corresponds to the power spectrum, and the second
    block to the VID. {\it Top}: Signal
    plus white noise corresponding to the COMAP1 experiment
    ($\sigma_\text{voxel} \approx 11 \, \mu$K). {\it Middle}: Signal
    plus white noise corresponding to the COMAP2 experiment
    ($\sigma_\text{voxel} \approx 8 \, \mu$K).  {\it Bottom}: Signal
    alone. Note that here we have changed the color scale. 
    {\it Left}: Covariance matrices without beam
    smoothing. {\it Right}: Covariance matrices with
    $\theta_\text{FWHM}=4'$ beam smoothing.\label{fig:cov}}
\end{figure*}

An advantage of using a Gaussian likelihood for the VID, is that 
it gives us a straightforward way to take into account the correlations 
between temperature bins apparent in the covariance matrix, $\xi_{ij}$
(e.g. in Figure \ref{fig:cov}). 
For another approach to building up a $\mathcal{P}(D)$ likelihood, see 
\cite{Glenn2010}.

To estimate $\langle d_i \rangle$, we compute 10 maps of the survey
volume at each step in the MCMC chain, using the current model
parameters $\theta$ with different dark matter halo realizations
(randomly drawn from 252 independent catalogs corresponding to the
survey volume). The specific number of
realizations, 10 in our case, represents a compromise between
minimizing the sample variance in the estimate of $\langle d_i
\rangle$ and maintaining a reasonable computational cost per MC step.
 Finally, we bin all the halos in the 10 realizations 
according to their luminosity, and use this histogram to estimate the
luminosity function at the current values of $\theta$. This way the
MCMC precedure gives us the luminosity function at different points
in parameter space, sampled according to the posterior of the model 
parameters, which we can use to derive constraints on the luminosity
function itself.

We adopt the same physically motivated priors as discussed by
\cite{Li}. Specifically, these read
\begin{align}
  P(\sigma_\mathrm{SFR}) &= \mathcal{N}(0.3,0.1)\\
  P(\log \delta_\text{MF}) &= \mathcal{N}(0.0,0.3)\\
  P(\alpha) &= \mathcal{N}(1.17,0.37) \\
  P(\beta) &= \mathcal{N}(0.21,3.74) \\
  P(\sigma_{L_{\mathrm{CO}}}) &= \mathcal{N}(0.3,0.1),
\end{align}
where $\mathcal{N}(\mu,\sigma)$ corresponds to a Gaussian distribution with 
mean $\mu$ and standard deviation $\sigma$.
Additionally, we require the
two logarithmic scatter parameters, $\sigma_\mathrm{SFR}$ and
$\sigma_{L_{\mathrm{CO}}}$, to be positive. We choose the mean of all these
distributions as the fiducial model, $\theta_0$.

To quantify the importance of joint PS+VID analysis, we perform the
above analysis both with each observable separately, and with the
joint analysis. The main result in this paper may then be formulated
in terms of the relative improvement on the CO luminosity function
uncertainty derived from the joint analysis to those found in the
independent analyses.

When calculating our observables (PS and VID), we assume that our survey
volume can be treated as a rectangular grid of voxels with constant
co-moving volume. We also neglect the evolution of our observables over
redshifts between $z = 2.4$ and $3.4$. That is, we assume that samples
from different redshifts are drawn from the same distribution, whether
they are power spectrum modes or voxel temperatures. We also assume that
the instrument beam is achromatic, and equal to the value at the 
central frequency. This is of course just an approximation that we make
in order for the analysis to be simple. If we were doing experiments with
higher signal to noise, we might divide our data into two different redshift 
regions and do an independent analysis of each region. This could allow us 
to study the redshift evolution of the observables. For COMAP (1 and 2),
however, we are probably best off combining all the data, like we do here, 
in order to increase the overall signal to noise. 

Finally, since COMAP will not measure absolute zero-levels, we
subtract the mean from all maps. For the power spectrum, this has a
negligible impact, as it simply removes one out of $N_\mathrm{vox}$ modes. However,
for the VID it has a significantly higher impact. Specifically, it
makes it much harder to distinguish a potential background of
weak sources from noise. Indeed, as shown by \cite{Breysse1}, removing
the monopole makes it much harder to detect a possible low luminosity
cutoff in the CO luminosity function using the VID.

\begin{figure*}[p]
\center
	\includegraphics[width=0.45\textwidth]{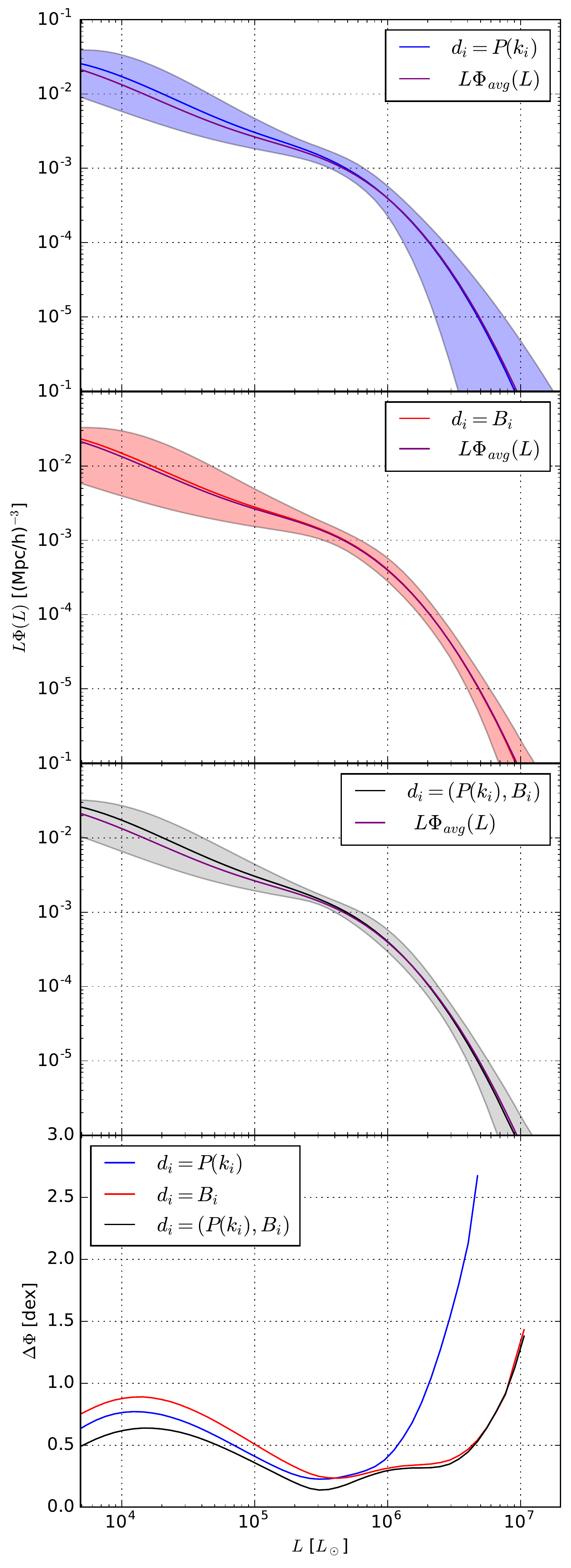}
	\includegraphics[width=0.45\textwidth]{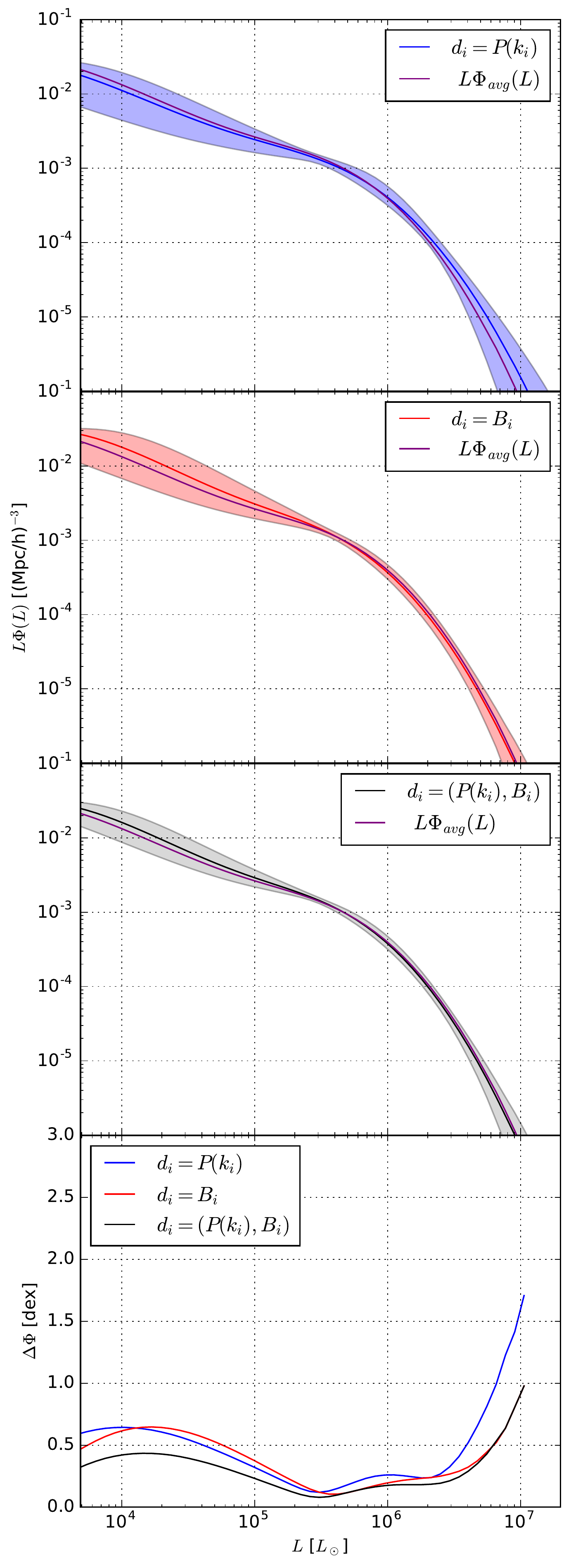}
  \caption{Constraints on the luminosity function from simulated experiments COMAP1 
  ({\it left}) and COMAP2 ({\it right}). The shaded area corresponds to 
  95\% credibility intervals, solid lines correspond to the median, 
  while the purple curve corresponds to the average luminosity function 
  derived from all the available halo catalogs (i.e. the ensemble mean). 
  {\it Top}: Constraints 
  derived using only the power spectrum $P(k_i)$ as the observable. 
  {\it 2.~row}: Constraints derived using only the temperature bin 
  counts $B_i$ as the observable. {\it 3.~row}: Constraints derived 
  by a joint analysis using both the power spectrum $P(k_i)$ and the 
  temperature bin counts $B_i$ as observables. \label{fig:lum1} 
  {\it Bottom}: Comparison of the uncertainty of the luminosity 
  function constraints in dex, i.e. $\Delta \Phi \equiv \log_{10} 
  \Phi_\mathrm{97.5 \%} - \log_{10} \Phi_\mathrm{2.5 \%}$.}
\end{figure*}

\begin{figure*}[p]
\center
	\includegraphics[width=1.0\textwidth]{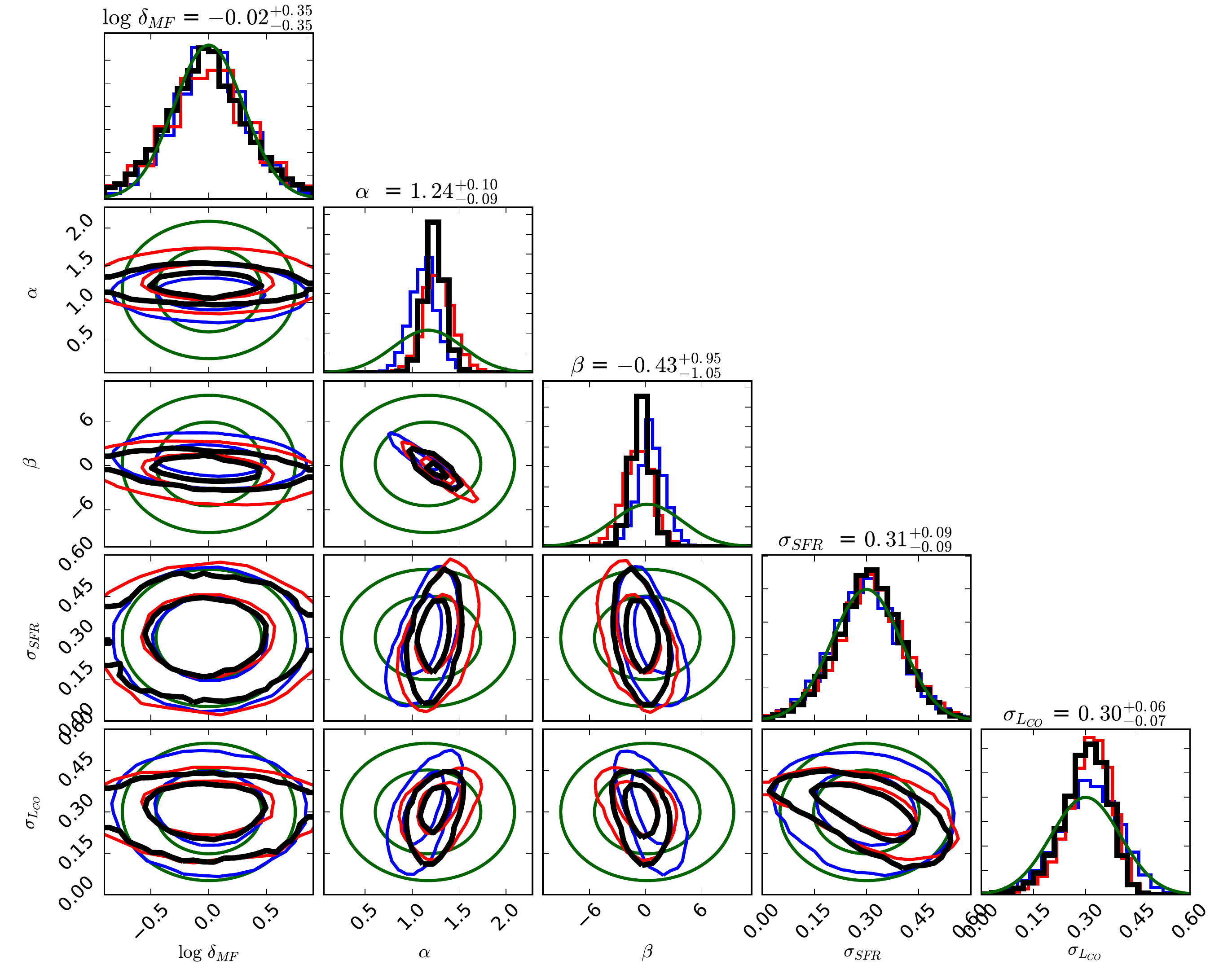}
  \caption{\label{fig:post}Posterior distributions for the \cite{Li} model parameters for 
  a single realization of the COMAP2 experiment (the same realization as 
  the COMAP2 results in Figure \ref{fig:lum1}). Results for PS, VID and 
  joint PS+VID analysis are shown in blue, red and (slightly bolder) black respectively. 
  Prior distributions are shown in green. The two curves of each color
  correspond 68\% and 95\% credibility regions. The numbers on top of each colum 
  correspond to the 68\% credibility interval for each parameter from the PS+VID 
  analysis. We see that while the 
  posterior of the two scatter parameters, $\sigma_\mathrm{SFR}$ and 
  $\sigma_{L_{\mathrm{CO}}}$ is mostly set by the prior, the posterior 
  on $\log \delta_\text{MF}$, from the SFR-$L_\mathrm{IR}$ relation, 
  is actually slightly wider than the prior, suggesting a significant 
  intrinsic scatter in estimates of this parameter. These results are consistent 
  with the corresponding results in Figure 7 in \cite{Li}. The two 
  parameters that are actually strongly constrained by the simulated 
  experiment are $\alpha$ and $\beta$, the two parameters from the 
  $L_{\mathrm{CO}}$-$L_{\mathrm{IR}}$ relation, and this figure shows 
  that, at least for this realization, the constraints on these two get 
  significantly improved in the combined analysis (PS+VID) as compared 
  to analysis using the individual observables. This figure was created 
  using the publicly available code\textsuperscript{a} \texttt{corner}
  \citep{corner}.}
  \raggedright\small\textsuperscript{a} \scalebox{.9}{https://github.com/dfm/corner.py}
\end{figure*}

\section{Results}

We are now ready to present the main numerical results from our
analysis, and we start with an inspection of the joint PS+VID
covariance matrix itself.

\subsection{Visual inspection of the PS+VID covariance matrix}

Figure~\ref{fig:cov} shows the pseudo-correlation matrices, $c_{ij}$,
for our two experimental setups, as well as for pure signal alone, for
reference. In order to illustrate the effect of the beam, we show
covariance matrices from maps both without and with beam smoothing in
the left and right columns, respectively.

The first thing to notice is that instrumental noise significantly
reduces the numerical values of the normalized covariance matrices,
bringing it closer to the independent white noise case for which
$c_{ij} = \delta_{ij}$. This agrees with intuition, since the noise
itself is white and uncorrelated. 

Beam smoothing also leads to weaker correlations. This is mainly due
to the beam diluting the signal at small scales, where the correlation
is otherwise strongest.

Next, we notice that the cross-correlations between the power spectrum
and VID are of the same order of magnitude as the correlations internal 
to each observable itself. Thus, it is essential to account for all these
correlations in any joint PS and VID analysis, as done in the present
paper. 

Finally, we note that when designing an experiment like COMAP, one of
the important trade-offs
involves observation time per
field. To obtain a fast signal detection it is in general advantageous
to observe deep on the smallest possible field. However, this only
holds true while the signal-to-noise per voxel is significantly less
than unity. When the noise starts to become comparable to the signal,
the signal-induced voxel-voxel correlations starts to become
important, and the effective uncertainties no longer scale as
$\mathcal{O}(1/\sqrt{\tau})$, where $\tau$ is the observation time per pixel. 
Generally, in such a tradeoff, any significant
correlations between different power spectrum modes or voxel
temperatures will tend to favor larger survey area or multiple
fields, both effectively leading to more independent samples, and
thereby higher overall integration efficiency.

\subsection{Luminosity function constraints}

We are now ready to present both individual and joint PS+VID
constraints on the CO luminosity function, and these are
summarized in Figure \ref{fig:lum1} for COMAP1 (\emph{left column}) and
COMAP2 (\emph{right column}). The top row shows the constraints
obtained from the power spectrum alone; the middle row shows the
constrains obtained from the VID alone; and the third row shows the
constraints from the joint analysis. In each panel, the shaded
colored region shows the 95\% credibility region from the MCMC
samples, and the solid line with the same color shows the posterior
median. The purple solid line shows the average luminosity function
obtained from the mean of all available halo catalogs, and thus
represents the ensemble average of our input model. Note that the
colored regions correspond to one single realization, and the
uncertainties therefore contain contributions from instrumental
noise, cosmic variance and sample variance. The agreement between the
estimated confidence regions and the ensemble mean is quite
satisfactory in all cases, with uncertainties that appear neither too
large nor too small.

Considering first the individual PS and VID estimates, shown in the
top two rows, we see that the two observables are indeed complementary. 
In particular, the VID primarily constrains the high
luminosity end of the luminosity function, while the power spectrum
imposes relatively stronger constraints on the low luminosity
end. This makes sense intuitively, since the VID is essentially
optimized to look for strong outliers above the noise, whereas the
power spectrum represents a weighted mean across the full field for
each physical scale. It is interesting to note, however, that the  VID
provides, on average, stronger constraints on the luminosity 
function than the power spectrum does.

Due to this complementarity,
the joint estimator provides the strongest constraints of all. 
To make this point more explicit, the fourth
row compares the uncertainties of the independent power
spectrum and VID analyses to the joint constraints. Of course, there
is a significant amount of cosmic variance in each of these functions,
and the precise numerical value of the uncertainty ratio therefore
varies significantly with luminosity; but the mean trend is clear: The
individual analyses typically result in 20--70\% larger uncertainties
than the joint analysis when averaged over luminosities between 
$L_\text{CO} = 10^4 - 10^7 L_\odot$. Over 10 cosmological realizations 
the PS and VID 
resulted in, on average, 58\% and 30\% larger uncertainties 
(in dex) individually, than the joint analysis. This is
the main novel result presented in this paper.

\subsection{Posterior distribution of model parameters}
Lastly we present the constraints of the model parameters themselves. 
When doing the MCMC posterior mapping we explore the parameter space of 
the \cite{Li} $L_\textrm{CO}(M_\textrm{Halo})$ model. Figure 
\ref{fig:post} shows the 
posterior distribution for these parameters derived from one realization 
of the COMAP2 experiment (the same realization as the COMAP2 results in 
Figure \ref{fig:lum1}). 

Results for PS, VID and joint PS+VID analysis are shown in blue, 
red and black respectively. Prior distributions are shown in 
green. The two curves of each color correspond 68\% and 95\% 
credibility regions.

\begin{figure}
\centering
	\includegraphics[width=0.5\textwidth]{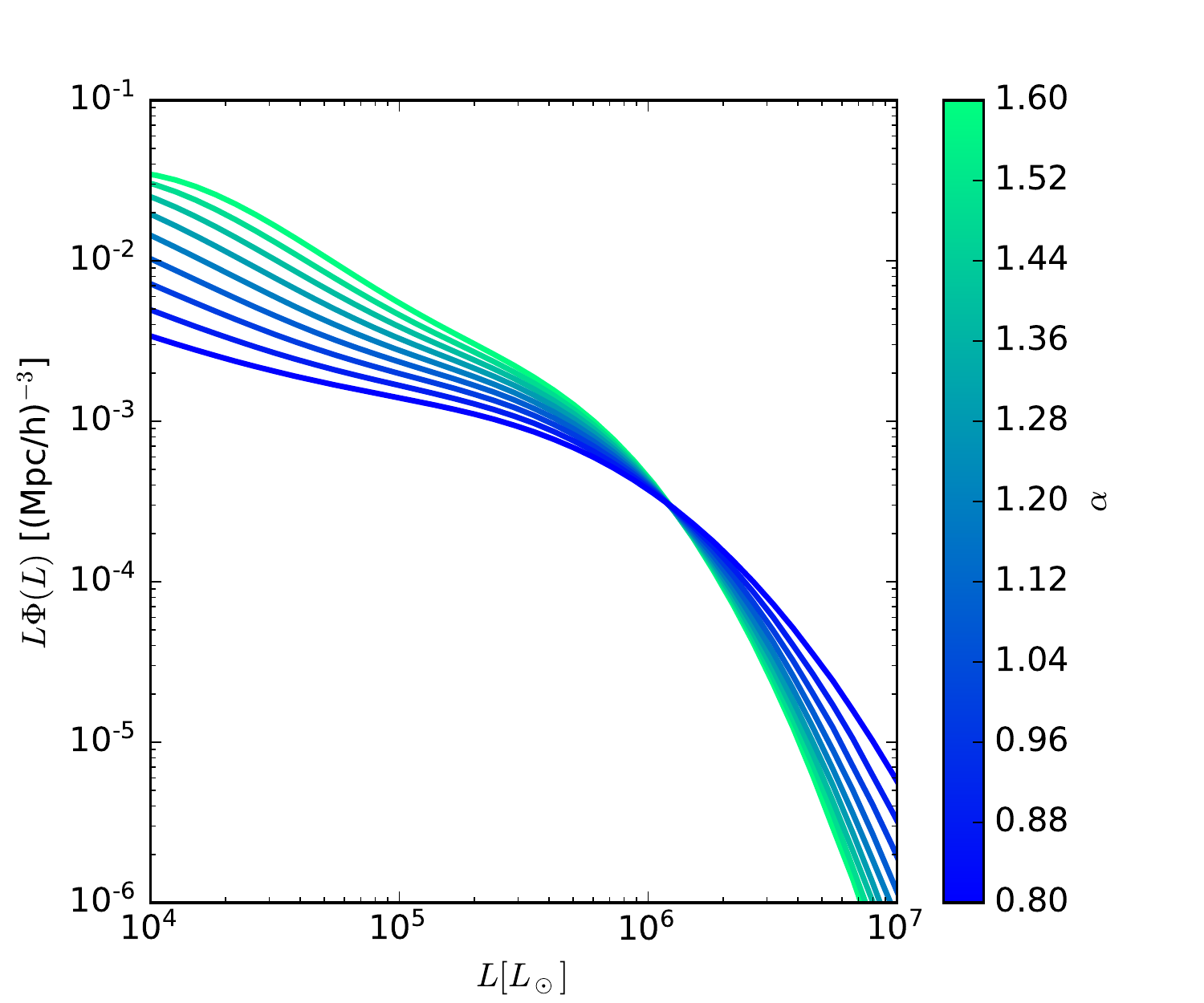}
  \caption{Plot of the CO luminosity function in the \cite{Li} model, 
  for different values of $\alpha$ and $\beta$. The colors of the lines
  indicate the values of $\alpha$, the values of $\sigma_\mathrm{SFR},$ $ 
  \sigma_\mathrm{L_\mathrm{CO}}$ and $\log \delta_\text{MF}$ are fixed
  at $0.3, 0.3$, and $0.0$ respectively, while the value of $\beta$ is 
  determined from the relation $\alpha = -0.1 \beta + 1.19$. This line
  corresponds roughly to the direction along which $\alpha$ and $\beta$ 
  are degenerate. Although the overall detectability of the signal
  remains roughly constant along this line, we see that the shape of
  the luminosity function changes significantly. We see that lower
  values of $\alpha$ correspond to less steep high luminosity tails
  in the luminosity function, meaning that a larger proportion 
  of the overall signal comes from high luminosity sources.
  \label{fig:lum_alpha}}
\end{figure}

We see that the two parameters that are mainly constrained are 
$\alpha$ and $\beta$, the two parameters from the average $L_{\mathrm{CO}}$-
$L_{\mathrm{IR}}$ relation.  These two parameters are fairly degenerate,
and the direction in which they are degenerate is given roughly by the line
$\alpha = -0.1\beta + 1.19$ \citep{Li}. In Figure \ref{fig:lum_alpha} we show the
luminosity function for different points on this line. For the figure,
the values of $\sigma_\mathrm{SFR},$ $ 
  \sigma_\mathrm{L_\mathrm{CO}}$ and $\log \delta_\text{MF}$ are fixed
  at $0.3, 0.3$, and $0.0$ respectively. Although the overall
  signal strength, at least in terms of detectability, is fairly constant 
  along this line, the shape of the luminosity function changes significantly.
  Lower values of alpha imply a more steep power law relation between 
  $L_{\mathrm{CO}}$ and $L_{\mathrm{IR}}$ leading to more sources with very 
  high or very low luminosity. We see this as a flattening of the luminosity
  function. In such a case a larger fraction of the overall signal will be
  given by high luminosity sources.

The other parameter that is also slightly constrained is the log-normal
scatter parameter from the $L_{\mathrm{CO}}$-$L_{\mathrm{IR}}$ relation, 
$\sigma_\mathrm{L_\mathrm{CO}}$. This parameter 
is only slightly more constrained compared to the prior, with the 
highest values of $\sigma_\mathrm{L_\mathrm{CO}}$ being disfavoured. 
The posterior of the other scatter parameter, $\sigma_\mathrm{SFR}$, is 
basically given by the corresponding prior (i.e. this parameter is not 
very well constrained by the experiment), although, as expected from
the fact that the scatter parameters have basically the same effect, 
we see signs of the degeneracy between them in the posterior. 

Interestingly, the normalization parameter in the 
SFR-$L_\mathrm{IR}$ relation, $\log \delta_\text{MF}$, actually has a
posterior that is wider than the prior. This may be because the 
best fit of this parameter from each of the different patches have an
intrinsic scatter larger than the scatter in the prior. 
We note that we see the same effect in 
\cite{Li} (their Figure 7). 

From the mean relations in the \cite{Li} model we have 
$\log L_{\mathrm{CO}} \sim - \beta - \log \delta_\text{MF}$.  
Intuitively, we would then expect $\log \delta_\text{MF}$ to be completely
degenerate with $\beta$. However, since the SFR-$L_\mathrm{IR}$ is
much better constrained by observations than the 
$L_{\mathrm{CO}}$-$L_{\mathrm{IR}}$ relation is, the prior 
on $\log \delta_\text{MF}$ is much tighter than the one on $\beta$.
The degeneracy thus prevents us from 
constraining $\log \delta_\text{MF}$ until $\beta$ is constrained
to a comparable level.

\section{Discussion}

We have developed a joint power spectrum and voxel
intensity distribution analysis for the CO luminosity function in the
context of the COMAP CO intensity mapping experiment. We have
implemented an efficient approach to estimating the joint covariance
matrix for these two observables, and shown that accounting for both
one- and two-point correlations leads to 20--70\% smaller uncertainties
on the CO luminosity function for both COMAP1 and COMAP2.

The critical computational engine in our approach is the construction
of fast yet semi-realistic simulations of the signal in
question. These simulations are based on the computationally cheap
peak patch dark matter halo simulations produced by \cite{bond1996, Stein2018}, coupled
to the semi-analytic CO luminosity model of \cite{Li}. Of course, the
results we derive are correspondingly limited by how well the model
reproduces the true cosmological signal. If the true signal is
significantly more complex than the model predicts, the constraints in
Figure \ref{fig:lum1} will not be reliable.

The strength of the constraints on the CO luminosity function will 
depend on the overall level of the CO signal, which is highly uncertain.
However, given the same rough level of signal, we expect the constraints on the 
luminosity function at the high luminosities
to be less model dependent than the constraints on the $L_{\mathrm{CO}} - M_\text{halo}$ 
relation or the luminosity function at lower luminosities. This is because 
the high luminosity sources leave a fairly unique imprint on the maps that
does not depend on the specific model used. 

Additionally, we expect that the relative merits of using the PS or the VID 
as observables will change depending on the properties of the signal. In 
particular, anything that increases the shot noise of the signal, like a 
a strong galactic duty cycle, a large intrinsic scatter in luminosities 
or just a more top-heavy luminosity function, will make the resulting map 
more non-Gaussian, tending to favor observables like the VID more as compared to the 
PS. We can see this effect directly in Figure \ref{fig:post}. 
The VID is better, compared to the PS, at ruling out low values of $\alpha$ and high values of 
$\sigma_{L_{\mathrm{CO}}}$, both of which corresponds 
to cases where we would have a more top-heavy luminosity function and thus
more shot noise. 

We also expect the map to be more non-Gaussian on small scales than on large,
so a wide survey with low resolution will tend to favor the PS, relative to the VID, 
more than a narrower high resolution survey.

 While the issues of model dependence are
less relevant for low signal-to-noise measurements, where we are just trying to
establish the rough level of the signal, they will
become more important as the measurements improve.

 Another potential issue with the simulations used in this 
paper is the minimum dark matter halo mass of 2.5$\times$10$^{10}M_\odot$. 
While the model used here predicts that only a small fraction of the CO signal
would come from halos lighter than this (see \cite{Li} and \cite{Chung2017}), 
other models could disagree. If fact, searching for a low luminosity cutoff
in the CO luminosity function is an interesting target for CO intensity 
mapping, and simulated halo catalogs with a smaller minimum dark matter halo 
mass would be useful both for forcasts and inference in such a scenario.

In general, it will be important to continuously improve
the simulation pipeline as the
experiment proceeds, in order to account for more and more 
cosmological, astrophysical and instrumental
effects. However, the most important point in our
approach is the fact that all such effects may be seamlessly accounted
for, as long as the simulation procedure is sufficiently fast in order
to be integrated into the MCMC procedure. 

It should also be noted that our approach may be generalized in many
different directions. For instance, the CO luminosity function does
not play any unique role in our analysis, but is rather simply one
specific worked example of a particularly interesting astrophysical
function to be constrained. Many other functions
may be constrained
in a fully analogous manner, including for instance non-parametric
$L_{\mathrm{CO}}(M_\text{halo})$ models, or any of the parameters that
are involved in converting the dark matter halo distributions to CO
luminosities.  The method is also not specific to CO 
intensity mapping, but should be equally well suited for other lines, 
or a combination of lines \citep{Chung2018}. 
Indeed, it should work for any type of random fields for which the covariance
matrix must be estimated by simulations. Finally, we also note that
there is nothing special about the power spectrum or VID as
observables, but any other efficient data compression can be equally
well included in the analysis, as long as the required compression
step is sufficiently computationally efficient.

\acknowledgments
Support for the COMAP instrument and operation comes through the NSF
cooperative agreement AST-1517598. Parts of this work were performed
at the Jet Propulsion Laboratory (JPL) and California Institute of
Technology, operating under a contract with the National Aeronautics
and Space Administration. HTI, HKE, MKF, and IKW acknowledges support
from the Research Council of Norway through grant 251328. 
Research in Canada is supported by NSERC and CIFAR. These calculations 
were performed on the GPC supercomputer at the SciNet HPC Consortium. 
SciNet is funded by: the Canada Foundation for Innovation under the 
auspices of Compute Canada; the Government of Ontario; Ontario Research 
Fund - Research Excellence; and the University of Toronto.
Work at Stanford University is supported by NSF AST-1517598 and by a 
seed grant from the Kavli Institute for Particle Astrophysics and Cosmology.
JOG acknowledges support from the Keck Institute for Space Studies, 
NSF AST-1517108, and the University of Miami.
HP's research is supported by the Tomalla Foundation.

We thank Sarah E. Church, Tim Pearson and other members of the COMAP 
collaboration for useful discussion and comments on a draft of this paper.

\bibliographystyle{aa}
\bibliography{vid_paper}

\end{document}